\documentstyle{elsart}
\input epsf.sty
\newcommand{\mincir}{\raise -2.truept\hbox{\rlap{\hbox{$\sim$}}\raise5.truept
\hbox{$<$}\ }}
\newcommand{\magcir}{\raise -2.truept\hbox{\rlap{\hbox{$\sim$}}\raise5.truept
\hbox{$>$}\ }}
\newcommand{\minmag}{\raise-2.truept\hbox{\rlap{\hbox{$<$}}\raise 6.truept\hbox
{$>$}\ }}
\newcommand{\be}{\begin{equation}}
\newcommand{\ee}{\end{equation}}
\newcommand{\ba}{\begin{eqnarray}}
\newcommand{\ea}{\end{eqnarray}}
\newcommand{\brr}{\begin{array}}
\newcommand{\err}{\end{array}}
\newcommand{\bc}{\begin{center}}
\newcommand{\ec}{\end{center}}

\newcommand{\bu}{\mbox{\bf u}}

\newcommand{\bq}{\mbox{\bf q}}
\newcommand{\bx}{\mbox{\bf x}}

\newcommand{\bk}{\mbox{\bf k}}

\newcommand{\hm}{\,h^{-1}{\rm Mpc}}
\newcommand{\vel}{\,{\rm km\,s^{-1}}}

\begin{document}
\begin{frontmatter}
\title{Cosmology using Cluster Internal Velocity Dispersions}
\author[PG]{S. Borgani},
\author[MI]{A. Gardini},
\author[TS]{M. Girardi} and
\author[AIP]{S. Gottl\"ober}
\address[PG]{INFN Sezione di Perugia, c/o Dipartimento di 
Fisica dell'Universit\`{a},
via A. Pascoli, I--06100 Perugia, Italy}
\address[MI]{Dipartimento di Fisica dell' Universit\`{a} di Milano,
    via Celoria 16, I--20133 Milano, Italy}
\address[TS]{Dipartimento di Astronomia, Universit\`a degli Studi di 
Trieste, Trieste, Italy; and SISSA via Beirut 2, I--34013 Trieste, Italy} 
\address[AIP]{Astrophysikalische Institut Potsdam, 
     An der Sternwarte 16, D14482 Potsdam, Germany}

\begin{abstract} 
We compare the cumulative distribution of internal velocity
dispersions of galaxy clusters, $N(>\sigma_v)$, for a large
observational sample to those obtained from a set of N--body simulations
that we run for seven {\sl COBE}--normalized cosmological scenarios. 
They are: the standard CDM (SCDM) and a tilted ($n=0.85$) CDM (TCDM) model,
a cold+hot DM (CHDM) model with $\Omega_\nu=0.25$, two low--density 
flat CDM ($\Lambda$CDM) models with
$\Omega_0=0.3$ and 0.5, two open CDM (OCDM) models with $\Omega_0=0.4$ and
0.6. The Hubble constant is chosen so that $t_0\simeq 13$ Gyrs in all
the models, while $\Omega_b=0.02\,h^2$ is assumed for the baryon
fraction. Clusters identified in the simulations are observed in
projection so as to reproduce the main observational biases of the
real data set. Clusters in the simulations are analysed by applying
the same algorithm for interlopers removal and velocity dispersion
estimate as for the reference observational sample. We find that
$\sigma_v$ for individual model clusters can be largely affected by
observational biases, especially for $\sigma_v\mincir 600\vel$. The
resulting effect of $N(>\sigma_v)$ is rather model dependent: models
in which clusters had less time to virialize show larger discrepancies
between intrinsic (3D) and projected distribution of velocity
dispersions. From the comparison with real clusters we find that both
SCDM and TCDM largely overproduce clusters. We verified for TCDM that
agreement with the observational $N(>\sigma_v)$ requires
$\sigma_8\simeq 0.5$. As for the CHDM model, it marginally
overproduces clusters and requires a somewhat larger $\sigma_8$ value
than a purely CDM model in order to produce the same cluster abundance. 
The $\Lambda$CDM model with $\Omega_0=0.3$ agrees with
data, while the open model with $\Omega_0=0.4$ and 0.6 underproduces
and marginally overproduces clusters, respectively.
\end{abstract}

\end{frontmatter}

\section{Introduction}
The abundance of clusters of galaxies has been recognized in the last
few years as a crucial constraint for cosmological scenarios of
large--scale structure formation (e.g., Bahcall \& Cen 1993; White,
Efstathiou \& Frenk 1993a); typical cluster masses, $\sim 5\times
10^{14}\Omega_0h^{-1}M_\odot$ ($H_0=100\,h\vel$Mpc$^{-1}$ is the
Hubble constant), corresponds to a scale length of about $10\hm$, so
that the number of clusters with such a mass determines the amplitude
of the fluctuation power spectrum on that scale, once the average
matter density is fixed. Rather simple analytical arguments, based on
the Press \& Schechter (1974, PS hereafter) approach to the
cosmological mass function, demonstrates that the cluster abundance is
exponentially sensitive to the power spectrum amplitude.  This allowed
several authors (e.g., Viana \& Liddle 1996; Eke, Cole \& Frenk 1996;
Pen 1996) to derive for CDM models tight relationships between the
r.m.s.  fluctuation within a top--hat sphere of $8\hm$ radius,
$\sigma_8$, and the density parameter $\Omega_0$. Although differing
in the details of the derivation, they converge to indicate that
$\sigma_8\Omega_0^{(0.5\!-\!0.7)} \simeq (0.5\!-\!0.6)$. Therefore,
once the value of $\Omega_0$ (and of the cosmological constant term)
is chosen, requiring at the same time to satisfy the $\sigma_8$
constraints, coming at small scales from the cluster abundance and at
large scales from measurements of CMB anisotropies (see, e.g., Bond \&
Jaffe 1996; Lineweaver et al. 1996), represents a stringent test for
the shape of the power spectrum. The most famous example is probably
represented by the standard CDM model, that, once normalized to match
the large--scale CMB anisotropies, overproduces clusters by at least
one order of magnitude.

It is however clear that, in order to fully exploit the potential of the 
cluster mass function as a cosmological test, one has to be sure about
{\em (a)} the reliability of the PS approach, and {\em (b)}
the accuracy of cluster mass measurements. 
As for the PS approach, its accuracy has been verified against a variety of
N--body simulations (e.g., White et al. 1993a; Lacey \& Cole 1994; Borgani et 
al. 1997a). As for cluster masses, methods for their determination are
based on virial analysis of internal galaxy velocities, X--ray temperature 
observations and  gravitational lensing of background galaxies. 
At present, the above three methods lead often to discordant
results for the same cluster (e.g., David, Jones \& Forman 1995,
Miralda--Escud\'e \& Babul 1995; Wu \& Fang 1997, and references therein).

Masses based on X-ray data are usually considered as the most reliable (e.g.,
Evrard, Metzler \& Navarro 1996, and references therein) and are widely 
employed to constrain cosmological models (see, e.g., Oukbir \& Blanchard 
1996, and references therein). However,
passing from temperature measurements to cluster masses relies on
assumptions, like spherical symmetry and hydrostatic equilibrium,
which may not be met in realistic cases. Observations of substructures
in the temperature patterns (e.g., Mohr et al. 1995; Henry \& Briel
1995; Buote \& Tsai 1996) and in the internal galaxy distribution
(e.g., Dressler \& Schechtman 1988; Escalera et al. 1994; Bird 1995;
Crone, Evrard \& Richstone 1996) indicates that clusters may not be
such completely relaxed structures, thus casting doubts on the
robustness of mass determinations based on hydrostatic equilibrium
(cf. Balland \& Blanchard 1996). Furthermore, the possible presence of
pressure supports of non--thermal origin, like intra--cluster magnetic
fields (e.g., Loeb \& Mao 1994; Ensslin et al. 1996) could also lead
to a mass underestimate.

As for gravitational lensing, there are at present only a few clusters 
at moderate or high redshift whose mass is estimated
from this method. Furthermore, weak lensing observations are more
reliable in providing the shape of the internal mass distribution, rather
than mass estimate (e.g. Squires \& Kaiser 1996; Seitz \& Schneider 1996).
Measurements of the overall cluster masses are based on strong lensing, 
whose application is however limited to very central regions.
In any case, all these estimates are rather dependent on the details of the
lens model (e.g. Bartelmann 1995).

Optical virial mass estimates have also been attempted by several authors
(e.g., Biviano et al. 1993). However, even in
this case, passing from the internal velocity dispersion of galaxies
to the cluster mass requires suitable assumptions about the degree of
virialization of the cluster, the nature of the galaxy orbits, the
relation between mass and galaxy distribution profiles (e.g., Merritt
1987) and, possibly, the presence of substructures (Bird 1995). 
A detailed analysis of biases affecting cluster mass estimates from
galaxy velocity dispersions have been recently pursued by Frenk et
al. (1996). From the analysis of numerical simulations they concluded
that virial methods can underestimate cluster masses by up to a factor
five.

Due to such uncertainties, a preferable procedure is to resort
directly on the line--of--sight velocity dispersion ($\sigma_v$,
hereafter), which is the observable quantity, instead of on the
cluster mass, as a diagnostic for cosmological models.  Cluster
velocity dispersions are rather easily obtainable from numerical
N--body simulations and, in fact, their distribution has been used as
a constraint for dark matter scenarios (Frenk et al. 1990; Bartlett \&
Silk 1993; Jing \& Fang 1994; Crone \& Geller 1994).

However, several observational biases may well affect $\sigma_v$
estimates, for instance connected to projection effects (e.g., White
et al. 1990; van Haarlem, Frenk \& White 1997) and to the limited
number of galaxies usually available to trace the internal cluster
dynamics (see, e.g., Mazure et al. 1996 and Fadda et al. 1996, for
recent detailed discussions about robust methods to estimate
$\sigma_v$). Moreover, cluster member assignment in projection,
anisotropy of galaxy orbits, cluster asphericity, infalling galaxies
and presence of substructures can significantly affect any measurement
of $\sigma_v$ and must be taken into account to reliably compare data
and simulations.

A further problem in the determination of the $\sigma_v$ distribution
concerns the availability of an observational sample which has to
satisfy well defined completeness criteria, to be implemented also in
the simulated sample. All the up-to-date available estimates of the
cumulative velocity dispersion function (CVDF hereafter),
$N(>\sigma_v)$, are based on samples that are complete with respect to
cluster richness (Girardi et al. 1993; Zabludoff et al. 1993; Collins
et al. 1995; Mazure et al.  1996, M96 hereafter). However, since the
relation between richness and $\sigma_v$ is intrinsically rather
broad, all samples that are complete in richness are not complete in
$\sigma_v$ (see Figure~8 of M96, see also Girardi et al. 1993).
Instead, they are systematically biased toward low $\sigma_v$'s.  For
instance, the ESO Nearby Cluster Survey (ENACS; Katgert et al. 1996),
which includes all the clusters with $R\ge 1$, is complete in velocity
dispersion only for $\sigma_v\magcir 800\vel$. In spite of all these
problems, there is a fair agreement among observational distributions
coming from different studies, at least within the completeness limits
of the corresponding samples (see, e.g., Fadda et al. 1996; F96
hereafter).  It is however clear that such observational limitations
must be taken into account in order to reliably compare real data to
simulations.

This paper is devoted to a close comparison of the $\sigma_v$
distribution from the sample analysed by F96 and from an extended set
of particle-particle--particle-mesh (P3M) simulations based on seven
different cosmological models. The original code on which the
simulations are based is the adaptive P3M one provided by Couchman
(1991).

As for the F96 sample, on which our analysis is based, it includes 153
Abell-ACO (Abell, Corwin \& Olowin 1989) clusters with redshift $z\le
0.15$. Data were taken from the published literature and from the ESO
Nearby Abell Clusters Survey (ENACS; Katgert et al. 1996).  All
clusters were selected so as to have at least 30 galaxies with
measured redshift within the cluster field and a significant peak in
redshift space ($>99\%$ c.l.; see F96 for a detailed description of
the sample). The inclusion of clusters of lower richness with respect
to ENACS pushes downwards the completeness limit for $\sigma_v$. As a
result, F96 estimate their CVDF to be complete above
$\sigma_{v,lim}\simeq 650\vel$.

Clusters extracted from the simulations are treated so as to reproduce
the main features of the observational analysis.  Firstly, they are
observed in projection with the same aperture radii and number of
galaxies with measured redshift as in the F96 sample. Then, we applied
an algorithm for removing interlopers and estimating $\sigma_v$ which
is the same as that of F96.

The aim of our analysis is twofold. From the one hand, the
availability in the simulations of the whole 3D information about the
cluster internal dynamics provides us with a test for the robustness
of the $N(>\sigma_v)$ estimate. From the other hand, the possibility
of analysing an extended set of simulations based, on different
initial spectra of fluctuations, allows us to put useful constraints on
the parameter space of cosmological models.

The plan of the paper is as follows.

We describe in Section 2 the simulations and the models that we
considered.  In Section 3 we outline the procedure for cluster
identification in 3D and how we simulate cluster observations to
reproduce the features of the F96 sample. After re-analysing the F96
sample, in Section 4 we show in detail how the algorithm, which
estimates $\sigma_v$ from projected data, works and discuss the
effects of observational biases on the $N(>\sigma_v)$ for the
different models. Afterwards, we compare the CVDF for data and
simulations and discuss the resulting constraints on cosmological
scenarios. A summary of the main results and our conclusions are
deserved to Section 5.

\section{Models and simulations}
As already mentioned the standard version of the CDM model (SCDM) turns out to
largely overproduce the abundance of galaxy clusters, once it is
normalized on large scales to match the measured CMB anisotropies.
Therefore, we decided to simulate SCDM, as a kind of reference
model, as well as six more models, which correspond to different
ways of improving SCDM. Such models can be divided into three main
categories as follows.
\begin{description}
\item[(a)] Two $\Omega_0=1$ models, namely a purely CDM model having a
``tilted" primordial spectral index (TCDM), $n=0.85$, and a Cold+Hot
Dark Matter (CHDM) model with $\Omega_\nu=0.25$ for the density parameter
contributed by one species of massive neutrinos with $m_\nu\simeq
6.25$eV.
\item[(b)] Two flat low--density CDM models, characterized by
$(\Omega_0,h)=(0.3,0.73)$ ($\Lambda$CDM$_{0.3}$) 
and $(\Omega_0,h)=(0.5,0.63)$
($\Lambda$CDM$_{0.5}$), respectively,
with the flatness provided by the cosmological constant term
$\Omega_\Lambda =1-\Omega_0$. For both models, the Hubble parameter $h$
is tuned in such a way to give $t_0\simeq 13$ Gyrs, so as to be
consistent with the $\Omega_0=1$ models.
\item[(c)] Two open CDM models, with $(\Omega_0,h)=(0.4,0.59)$
(OCDM$_{0.4}$), and $(\Omega_0,h)=(0.6,0.55)$ (OCDM$_{0.6}$), with the
same criterion as before for the choice of the Hubble parameter.
\end{description}
A further way to suppress cluster abundance in CDM models, that we do
not explore here, would be to abandon scenario of random phase
adiabatic fluctuations. For instance, a $\Omega_0=1$ CDM cosmology
with topological defects can easily provide $\sigma_8\simeq 0.5$
(e.g., Pen, Spergel \& Turok 1994).

For the CDM models we adopted the transfer function by Bardeen et al.
(1986), with shape parameter $\Gamma = \Omega_0h
\exp(-\Omega_b-(h/0.5)^{1/2}\Omega_b/\Omega_0)$, corrected to account
for a non negligible baryon contribution (Sugiyama 1995). The baryon
fraction has been chosen to be $\Omega_b=0.02 h^{-2}$ for all the
models. This value corresponds to the 95$\%$ upper limit from
primordial nucleosynthesis predictions (see, e.g., Copi, Schramm \&
Turner 1995; see Burles \& Tytler 1996, for higher $\Omega_b$ from
observations of low deuterium abundance in high redshift systems; see,
however, Fields et al.  1996 for a lower $\Omega_b$ prediction).  As
for the CHDM power spectrum, it has been explicitely computed by
following the linear evolution of the matter and radiation fluids
through the epoch of matter--radiation equality and recombination
epochs, down to the redshifts relevant for large--scale structure
formation.  We normalized all the spectra to match the four year {\sl
  COBE} data (e.g., Bennett et al. 1996), following the recipe
provided by White \& Scott (1996).  The resulting values for the
r.m.s. fluctuation amplitude within a top--hat sphere of $8\hm$
radius, $\sigma_8$, are reported in Table 1, where also the model
parameters are specified.

Also reported in the last column of Table 1 are the $\sigma_8$ values
predicted by the Eke et al. (1996) [E96 hereafter; see their eqs.
(6.1) and (6.2)] to reproduce the distribution of X--ray cluster
temperatures by Henry \& Arnaud (1991), based on the Press \&
Schechter (1974) approach and the assumptions of isothermal gas
distribution and hydrostatic equilibrium. Other approaches have been
derived by different authors, who provided slightly different results.
For instance, Viana \& Liddle (1996) found that $\sigma_8\simeq 0.6$
is required for $\Omega_0=1$ when normalizing the temperature function
only at $T_X=7$ keV.  Pen (1996) pointed out that the E96 scaling
provides too small $\sigma_8$ values if $\Omega_0<1$. For
$\Omega_\Lambda =0.65$ he found the $\sigma_8$ value by E96 to be
underestimated by about $17\%$. 
In general, such ocnstraints from X--ray data agree with the cluster
abundance as inferred, although with larger uncertainties, from the
frequency of large--separation lenses (see, e.g., Kochanek 1995).
Therefore, we regard the numbers reported in the last column of Table
1 more as guidlines than as stringent constraints.  Having this in
mind, we note that our models are not in general tuned so as to
reproduce such predictions. Indeed, our purpose here is to verify
whether a careful analysis of cluster velocity dispersions leads to
the same conclusions as the above analytical approaches, 
rather than picking up the best--fitting cosmological scenario.

The reported $\sigma_8$ value for TCDM assumes a vanishing tensor
(gravitational wave) contribution to the CMB anisotropy. In addition
to this, we also consider two more outputs at $\sigma_8=0.67$ and
$\sigma_8=0.51$. The first value corresponds to assuming $T/S=7(1-n)$
for the ratio between tensor and scalar contributions to the CMB
anisotropy, as predictied by power--law inflation (see, e.g.,
Crittenden et al. 1993). The second value is consistent with the
observational constraint reported in column 7.
 
\begin{table}[tp]
\centering
\caption[]{The model parameters.
Column 2: the density parameter $\Omega_0$; Column 3:
the cosmological constant term $\Omega_{\Lambda}$;
Column 4: the Hubble parameter $h$; Column 5: the 4 year {\sl COBE} predicted
linear r.m.s. fluctuation amplitude at $8\hm$ $\sigma_8$; Column 6: the 
initial redshift; Column 7: The $\sigma_8$ value predicted by Eke et al.
(1996; see text).}
\tabcolsep 5pt
\begin{tabular}{lcccccc} \\ \\ \hline \hline
 Model & $\Omega_0$ & $\Omega_{\Lambda}$ & $h$ & $\sigma_{8,COBE}$ & $z_i$ &
$\sigma_{8,E96}$ \\ \hline
SCDM & 1.0 & 0.0 & 0.50 & 1.20 & 35 & $0.52\pm 0.04$ \\
TCDM & 1.0 & 0.0 & 0.50 & 0.85 & 25 & $0.52\pm 0.04$ \\
CHDM & 1.0 & 0.0 & 0.50 & 0.77 & 20 & $0.52\pm 0.04$ \\ \\

$\Lambda$CDM$_{0.3}$ & 0.3 & 0.7 & 0.73 & 1.10 & 40 & $0.94\pm 0.07$ \\
$\Lambda$CDM$_{0.5}$ & 0.5 & 0.5 & 0.63 & 1.22 & 40 & $0.73\pm 0.06$ \\ \\

OCDM$_{0.4}$ & 0.4 & 0.0 & 0.59 & 0.58 & 35 & $0.77\pm 0.06$ \\
OCDM$_{0.6}$ & 0.6 & 0.0 & 0.55 & 0.90 & 35 & $0.64\pm 0.05$ \\ \\ \hline

\end{tabular}
\label{t:dm}
\end{table}

We simulate the development of non--linear gravitational clustering by
resorting to the adaptive P3M code developed by Couchman (1991). For each
model we run three realizations within a box of size $L=150\hm$, while
only one realization is run for CHDM. The evolution
of the density field is traced by following the trajectories of $128^3$ 
cold particles. In additions to these, for the CHDM model we also put
$2\times 128^3$ hot particles, in order to sample the neutrino
phase--space. Therefore, for all the purely CDM models the particle mass
is $4.5\times 10^{11}\Omega_0h^{-1}M_\odot$, while cold and hot particle
masses for the CHDM simulation are about $3.4\times 10^{11}h^{-1}M_\odot$
and $5.6\times 10^{10}h^{-1}M_\odot$, respectively.
Therefore, a typical cluster of mass $\simeq 10^{14}h^{-1}M_\odot$ is
resolved with more that $200\,\Omega_0^{-1}$ particles, thus ensuring an
adequate mass resolution.  

Initial conditions are realized by generating a random realization of 
the linear gravitational potential on a $128^3$ grid. Particles are moved
from their initial grid position ${\bf q}$ according to the Zel'dovich
approximation $\bx=\bq -\sum_{\bk}\bk \phi_k e^{-i\bx\cdot \bq}$
with comoving peculiar velocity
$\bu =\dot{\bx} = -\dot{a} 
\sum_{\bk} \alpha_k \bk \phi_{\bk} e^{-i\bx\cdot \bq}$,
where $\alpha_k=d\log \delta_{\bk} /d\log a$. Accordingly,
$\alpha_k={\rm const}$ for purely CDM models ($\alpha_k=1$ if
$\Omega_0=1$), while it has to be explicitely computed for the CHDM model
in order to account for the effect of residual neutrino free streaming on
the fluctuation growth (see also Ma 1996; Klypin, Nolthenius \& Primack
1996).

As for the hot particles in the CHDM simulation, in addition to the
gravitational velocities, they were given also a
thermal velocity which is randomly taken from the Fermi--Dirac
distribution
\be
f_{FD}(v)\,\propto \,{v^2 \over \exp\left[m_\nu v/k_B T_\nu(z_i)\right]+1}
\,.
\label{eq:fd}
\ee
Here $m_\nu\simeq 6.25$eV is the neutrino mass corresponding to
$\Omega_\nu=0.25$, $k_B$ is the Boltzman constant and
$T_\nu(z_i)=(1+z_i)T_{\nu,0}$, being $T_{\nu,0}=(4/11)^{1/3}T_{CMB,0}\simeq
1.95$ K the present--day temperature of the relic neutrino background.
Each pair of hot particles, which are initially located on the same grid
position, are assigned thermal velocities having the same modulus but
opposite directions, so as to ensure local momentum conservation. 

As for the long range force, it is computed on a $128^3$ grid, while the
short--range force is softened at scales smaller than
$\epsilon=0.1(L/128)$. Simulations are started from an initial redshift
$z_i$ corresponding to the epoch at which $\sigma=0.2$ for the linearly
estimated r.m.s. fluctuation on the grid. The resulting values of $z_i$ are
listed in column 6 of Table 1.   The integration variable has been taken
to be $p=a^{3/2}$, being $a=(1+z)/(1+z_i)$ the expansion factor. We
adopted a constant step size $\Delta p \simeq 0.35$ for all the models,
except for CHDM, for which we take $\Delta p\simeq 0.3$ to allow for a
slightly more accurate integration. 
Thanks to their dynamical and mass resolution, our simulations
are well suited to follow the internal dynamics of galaxy
clusters on the scales relevant for the $\sigma_v$ estimate.

\section{Construction of the cluster samples}

\subsection{Cluster identification in 3D}
Our method to identify clusters in the simulation box is based on the 
friend--of--friend (FOF) algorithm. The list of candidate clusters is
constructed by finding groups of CDM particles using a linking length $b=0.2$
(in units of the mean particle separation; see, e.g., Frenk et al. 1990; 
Lacey \& Cole 1994). In order to approach a more observational
selection procedure, for each FOF group we estimate the center of mass 
and draw around it a sphere having radius equal to the Abell one, 
$r_{Ab}=1.5\hm$. The center of mass of all the cold particles falling 
within this sphere is then computed and used as the starting point for the
next iteration. The sphere is moved around until we get convergence for 
its mass and position. We always find that few ($\sim 5$) iterations are
required to obtain a stable result. 
Cold particles are used to trace their internal velocity dispersion.
When the distance between the centers of two clusters is smaller than 
$2r_{Ab}$, the less massive cluster is removed from the list. We find that
this situation occurs in most cases when a small group is located at 
the outskirts of a massive object.  

For the above choice of the FOF linking length, the group--finding
algorithm picks up structures with $\rho/\bar \rho \simeq
180\,\Omega_0$ for the average internal overdensity, thus very close
to the virialization overdensity for $\Omega_0=1$. If $\Omega_0 < 1$,
smaller $b$ values are required to pick up virialized structures (see,
e.g., E96); it turns out that $b\simeq 0.15$ is required in the most
extreme case of the $\Lambda$CDM$_{0.3}$ model.  However, since FOF
groups are only used as initial points from where to start looking for
clusters, one expects the final cluster list not to be very sensitive
to the choice of $b$. We verified that final results are insensitive
to variations of $b$ in the range 0.15--0.2.

\subsection{Observing simulated clusters}
The list of clusters identified in 3D is used as the starting point to 
obtain a sample which resembles as close as possible the features of realistic 
cluster observations. The procedure to pass from the 3D to the projected 
estimate of the cluster velocity dispersion is realized according to the 
following steps.
\begin{description}

\item[(1)] Each 3D cluster is observed by selecting all the particles
contained within a cylinder of fixed aperture radius, $r_a=1.5\hm$, which
extends for $\pm 8,000\vel$ in line--of--sight velocity from the cluster 
center,
by allowing for periodi boundary conditions.
Each cluster is observed three times along the directions of the three
coordinate axes.  Galaxy velocity distributions for observed clusters
do not extend in general outside $\pm 4000$ km s$^{-1}$ from the mean
cluster velocity (see, e.g., Zabludoff, Huchra \& Geller 1990).
Hence, the extension of $\pm 8000 \vel$ in l.o.s of our selected
cylinders is sufficiently large to allow the density peak finding
algorithm to retrieve cluster peaks. In particular, we avoid spurious
effects at the border of the velocity range, that may occur when a
peak of a close cluster is sharply truncated there. We verified that
this happens few times if a smaller extension of $\pm 4000 \vel$ is
considered.

\item[(2)] In the observational case the sampling density of cluster
galaxies is by far much smaller than that allowed in the
simulations and improves with cluster richness (as we verified by using
the data sample analysed by F96).  Furthermore, an aperture radius as
large as $1.5 \hm$ is in general appropriate only for very rich
clusters (see, e.g., Katgert et al. 1996), poor clusters and groups
requiring $r_{a}\simeq 0.5\hm$ (e.g., Dell'Antonio, Geller \&
Fabricant 1995).  Therefore, observing the latter at a much larger
aperture one may include a nearby close cluster, whose presence would
pollute the velocity dispersion estimate. To account for this, we group
clusters of the observational sample (F96) into four equally spaced
bins in $\sigma_v$, from 0 to $1300\vel$. Then, we assign to a model cluster
with velocity dispersion $\sigma_v$ the
aperture radius $r_a$ and the number of galaxies $N_g$ of a 
real cluster which is randomly selected between those belonging to 
the same $\sigma_v$ bin. In this
way, we generate a $r_a$--$\sigma_v$ and a $N_g$--$\sigma_v$
correlation, having the same shape and dispersion as for the
observational sample.

\item[(3)] Galaxy membership to clusters is assigned by following the
same procedure of F96. 

Starting from the discrete distribution of galaxies in velocity space
within each cylinder, a continuous distribution is obtained by
applying the adaptive kernel method (see Pisani 1993, 1996 and
references therein). This method is based on convolving the discrete
distribution with a Gaussian kernel, whose r.m.s. amplitude is chosen
point by point so as to minimize the difference between the original
distribution and its reconstructed continuous representation. Peaks of
this continuous distribution which are significant at $\ge 99\%$ level
are then identified with clusters.  
In the case of two or more peaks,
we retain only the one closer to the 3D cluster position on which the
cylinder is centered, so as to avoid multiple counting of the same
object.

Then we applied the ``shifting gapper" method which is an iterative
procedure based on combining projected position and velocity
information for each galaxy.  A galaxy is considered as interloper if
it is separated by more than 1000 $\vel$ from the central body of the
velocity distribution of galaxies lying at the same distance (within a
bin of 0.4 $\hm$ width) from the cluster center. Therefore, this method
accounts for the possibility that the velocity dispersion strongly
depends on the clustercentric distance, as seen in several cases for
observed clusters (see, e.g., F96 and Girardi et al. 1996, G96
hereafter).

Once the cluster membership has been assigned, the corresponding
$\sigma_v$ is estimated by using the robust estimator described by Beers,
Flynn \& Gebhardt (1990).

As in F96, we discarded from the sample those clusters whose
$\sigma_v$ has a bootstrap error larger than $150\vel$. A
visual inspection of the velocity dispersion profile of such clusters
systematically reveals that these large errors occur when $\sigma_v$
widely oscillates at large radii, instead of smoothly converging to a
constant value. This could be the signature for an insufficient sampling 
or for the presence of a
heavy contamination due to unremoved interlopers, whose effect is
that of boosting $\sigma_v$ for an otherwise low velocity dispersion
group.
\end{description}

Although this procedure is designed so as to reproduce the
observational situation as close as possible, it is clear that several
aspects have not been included, due to intrinsic limitations of our
simulations. In the following we discuss these limitations and comment
their effects on the final results.
\begin{description}

\item[(a)] The projected richness of optically selected clusters can
be in general a bad predictor of the real 3D richness (see, e.g.,
Frenk et al.  1990). For instance, van Haarlem et al. (1997) pointed out that
{\em (i)} about one third of $R\ge 1$ Abell clusters identified in
$\Omega_0=1$ CDM N--body simulations can arise from superposition of
intrinsically poor clusters, and {\em (ii)} 30\% of intrinsically rich
clusters are missed and classified as poor clusters because of
fluctuations of the background galaxy counts. On the contrary, Cen
(1996) addressed the same issue by analysing an $\Omega_0=0.4$ open
CDM simulation and argued against a substantial richness contamination
by projection. Whether such a disagreement is due to the different
nature of the simulated models (in an open Universe clusters are
smaller, more concentrated structures than for $\Omega_0=1$) or to
differences in the simulations, we regard this as an open question.
As far as our comparison with observational data is concerned, we
point out that misclassified clusters ought not to represent a serious
problem. In fact, even though a cluster can be misclassified in
projection, it can be recognized as a superposition of poor clusters
when observed in redshift space (e.g. in the case of A151 and A367,
see F96). As for the missed rich clusters, M96 showed that the effect
of this bias is that of reducing the range of completeness of the
observed $\sigma_v$ distribution thanks to the existence of a 
broad, but well defined, correlation between the cluster richness
in two and three dimensions (see Figure~7 in
their paper).  Since our reference CVDF
is estimated to be complete down to $\sigma_{v,lim}\simeq 650$
km s$^{-1}$ (see F96) we account for this bias by comparing
$N(>\sigma_v)$ for data and simulations only at $\sigma_v \ge
\sigma_{v,lim}$ (see \S~4.2 below). This avoids any need to define the
richness for model clusters, which necessarily passes through a
problematic galaxy identification in the simulations.

\item[(b)] In their analysis, F96 removed the contribution of 
late--type galaxies, which is expected to be more affected by interlopers, 
when their velocity dispersion is significantly larger than that of 
early--type galaxies.
However, since any distinction between early-- and late--type galaxies is not 
allowed in our simulations, the analysis of the observational sample has been
repeated by including galaxy of any morphology (see \S~4.1). 
We expect this to turn into
an increase of the $\sigma_v$ estimates for the combined effects of a 
larger number of interlopers and of the larger velocity dispersion of 
spirals with respect to ellipticals (cf. also Stein 1996). 

Furthermore, F96
also attempted to remove the contribution of velocity gradients,
possibly produced by cluster substructures, cluster rotation or by the 
interaction with surrounding structures, like filaments and nearby clusters.
Since such effects are already accounted for in the simulations, 
we have chosen not to correct for them in the reanalysis of the 
observational sample (see \S 4.1). 

\item[(c)] Our comparison between data and simulations assumes that
galaxies are fair tracers of the internal cluster dynamics, that is,
no velocity bias is introduced. Whether this represents a reasonable
approximation is a matter of debate. Carlberg (1994)
found in its purely gravitational simulation that galaxy halos suffer
for a substantial velocity bias. This result has been confirmed by
Evrard, Summers \& Davis (1994) and by Frenk et al. (1996), who also included gas dynamics. On
the contrary, Katz, Hernquist \& Weinberg (1995) and Navarro, Frenk \&
White (1995) came to a different conclusion, claiming for no
substantial velocity bias in their cluster simulations. (Note that the
mass resolution of our simulation does not allow to address this
issue.)  As for observations, comparisons of galaxy velocity
dispersions and X--ray temperatures in clusters (Edge \& Stewart 1991;
Lubin \& Bahcall 1993; G96; Lubin et al. 1996)
indicate that the former provide a fair tracer of the dark matter
velocity dispersion (see, however, Bird, Mushotzki \& Metzler
1995). In the following we will show explicitely only for the SCDM 
simulations the effect of introducing the velocity bias found by
Evrard et al. (1994) for their cluster simulations based on this same model.
\end{description}

\section{Results}
\subsection{The Observational Distribution of Velocity Dispersions}

In order to allow for a more homogenous comparison with simulation
results (see point {\bf (b)} of \S 3.2), we used as a starting point 
the F96 sample in an
intermediate step of its analysis, that is before applying
removing late--type spirals and correcting for velocity gradients
(see \S~2.2 and \S~3 of F96). Moreover,
we considered only galaxies observed within a maximum aperture radius
of $1.5\hm$. In the case of multipeaked clusters (see \S~4 in F96), we
selected only the most significant peak.  We computed $\sigma_v$ for
each cluster and the resulting CVDF 
by following the same procedure as F96. In particular,
we weighted each cluster according to the richness--class distribution
of the Edinburgh--Durham Cluster Catalogue (EDCC; Lumsden et
al. 1992) in order to account for the
volume--incompleteness of the observational sample (see \S~5 of F96).

\begin{figure}
\mbox{\epsfxsize=14 cm\epsffile{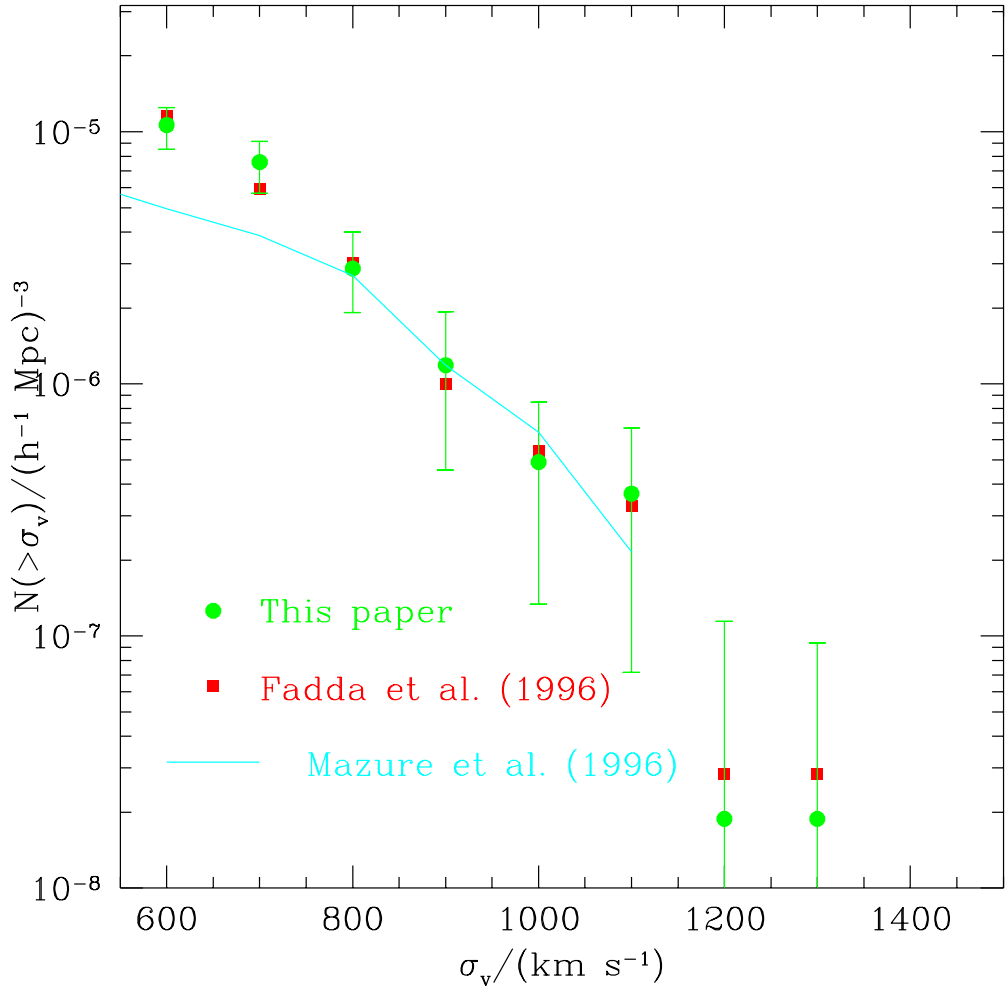}}
\caption{The cumulative velocity dispersion function for the 
observational sample as originally published by F96 (squares) and 
re-estimated here (circles).
The continuous line represents the determination by Mazure et al. (1996) for
ENACS clusters. Errorbars are
the 1$\sigma$ uncertainties obtained by adding in quadrature errors coming 
from the Poissonian statistics and from the bootstrap error in the 
determination of individual cluster velocity dispersions.}
\label{fi:F96}
\end{figure}

We compare in Figure 1 the original $N(>\sigma_v)$ from F96 (squares)
to the one we calculated here (circles).  Also plotted is the CVDF
obtained by M96 for the ENACS clusters (continuous line).  As already
pointed out in the Introduction, the ENACS sample is incomplete for
$\sigma_v\mincir 800\vel$. This explains the bending of its
$N(>\sigma_v)$ in this velocity range. As for the F96 sample, it is
remarkable the agreement with the original $N(>\sigma_v)$ estimate.
This indicates that, at least from a statistical point of view, 
velocity gradients and galaxy morphology play a
marginal role. In turn, these results are very close to those by M96
in the range of completeness of the ENACS.  The plotted errorbars
(reported only in one case for sake of clarity) represent the
$1\sigma$ uncertainties obtained by summing in quadrature the
Poissonian errors and the bootstrap errors in the determination of
$\sigma_v$ for each individual cluster.  Note that Poissonian errors
are due to the fact that we sample the intrinsic $N(>\sigma_v)$
distribution with a finite number of clusters, while bootstrap errors
are connected to the finite sampling of the velocity dispersion within
individual clusters. Therefore, such errors have independent origin
and must both be considered.

In the following, our CVDF will be used for the comparison 
with simulation results.

\subsection{Testing the method}
We will test here the reliability of the method for cluster 
membership assignment, that we described in the previous section.
We will also compare the resulting $\sigma_v$ estimates to the
intrinsic ones, as provided by using the whole 3D information from the
simulation particle distribution. After analysing how projection and sampling 
biases affect the determination of the velocity dispersion, we will also show
their overall imprint on $N(>\sigma_v)$.

We plot in Figure 2 
the result of introducing observational biases on four clusters 
selected from the TCDM model with $\sigma_8=0.67$. The first three clusters 
from the left have progressively larger $\sigma_v$ values, while the 
fourth column shows the case of a cluster which is rejected on the ground 
of its too large $\sigma_v$ bootstrap error [$>150\vel$; cf. point {\bf (3)}
in the previous section]. Panels in the first line show how the clusters 
appear in projection, with open circles indicating the rejected interlopers
and filled circles for the ``galaxies'' recognized as genuine members. We 
note that the algorithm for membership assignment is rather efficient
in identifying interlopers preferentially at the outskirts of the clusters,
while genuine members are more concentrated to define the cluster shape.
This is also confirmed by the panels in the second line, which show the
real--space distribution along the line of sight for accepted 
(filled histogram) and rejected (open histogram) members. In general, true 
members of the first three clusters are correctly recognized to lie
very close to the cluster center, with few exceptions (e.g., note the group
of seven galaxies, located at $\simeq 20 \hm$ from the center of the 
second cluster). As for the fourth cluster, a non--negligible fraction of 
galaxies recognized as genuine members are instead interlopers, which lie at
a distance $\magcir 20\hm$ from the cluster.

\begin{figure}
\mbox{\epsfxsize=14 cm\epsffile{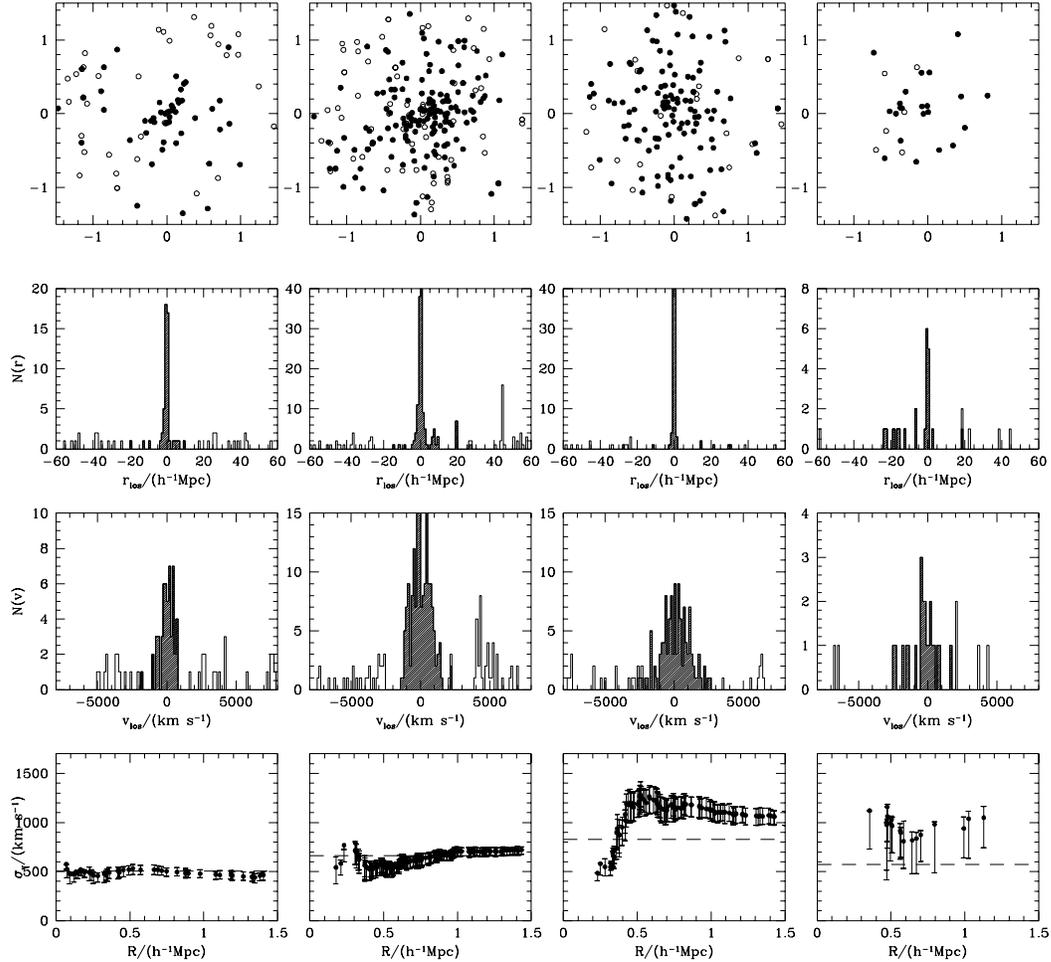}}
\caption{Results of the application of the cluster member assignment
and $\sigma_v$ estimate on four clusters extracted from the TCDM simulations
with $\sigma_8=0.67$. Panels in the first line show clusters in projection. 
Filled and open circles refer to ``galaxies'' recognized as genuine members 
and interlopers, respectively. Panels in the second and third lines show
the histograms for the line--of--sight galaxy distribution toward the cluster 
centers in real and velocity space, respectively. Open and filled histograms 
are for the distributions of interlopers and true cluster members. Panels in 
the fourth line are for the integral velocity dispersion profiles. The 
horizontal dashed line corresponds to the intrinsic $\sigma_v$ value, 
computed at $1.5\hm$. Errorbars are for the r.m.s. scatter over 1000 
bootstrap resamplings. For each cluster the point at the smallest 
radius represents $\sigma_v$ as computed for the first six innermost 
galaxies.}
\label{fi:test}
\end{figure}

Panels in the third line show the redshift--space distribution from which
$\sigma_v$ is actually calculated, while the fourth line shows the projected
integrated 
velocity dispersion profile. In each panel the horizontal dashed line 
represents the intrinsic value of $\sigma_v$, which is estimated from all the
simulation particles lying within $1.5\hm$ from the cluster center. Errorbars
are the $1\sigma$ scatter from 1000 bootstrap resamplings. As for the 
first two clusters, it is remarkable how well the correct $\sigma_v$ is 
recovered at the scales where it keeps flat (see F96 and G96 for
discussions about the flatness of profiles
in the external cluster region). As for the third cluster, 
despite the fact that the $\sigma_v$ profile flattens at scales 
$\magcir 1\hm$, it does not converge to the correct value. Note that the
discrepancy of $\sim 200\vel$ exists at a high confidence level ($>3\sigma$)
and, therefore, can not be ascribed to sampling uncertainties. This means
that a better recovering of the intrinsic $\sigma_v$ can hardly be attained 
in this case with a denser galaxy sampling.
The situation is quite different for the fourth 
cluster. In this case, the discrepancy is as large as $\sim 400\vel$, but 
it is not significant, as a consequence of the large errorbars. This example 
shows the effectiveness of eliminating those clusters with large $\sigma_v$ 
errors, which in general corresponds to rather small and loose structures,
whose $\sigma_v$ can be heavely boosted by sampling uncertainties.

\begin{figure}
\mbox{\epsfxsize=14 cm\epsffile{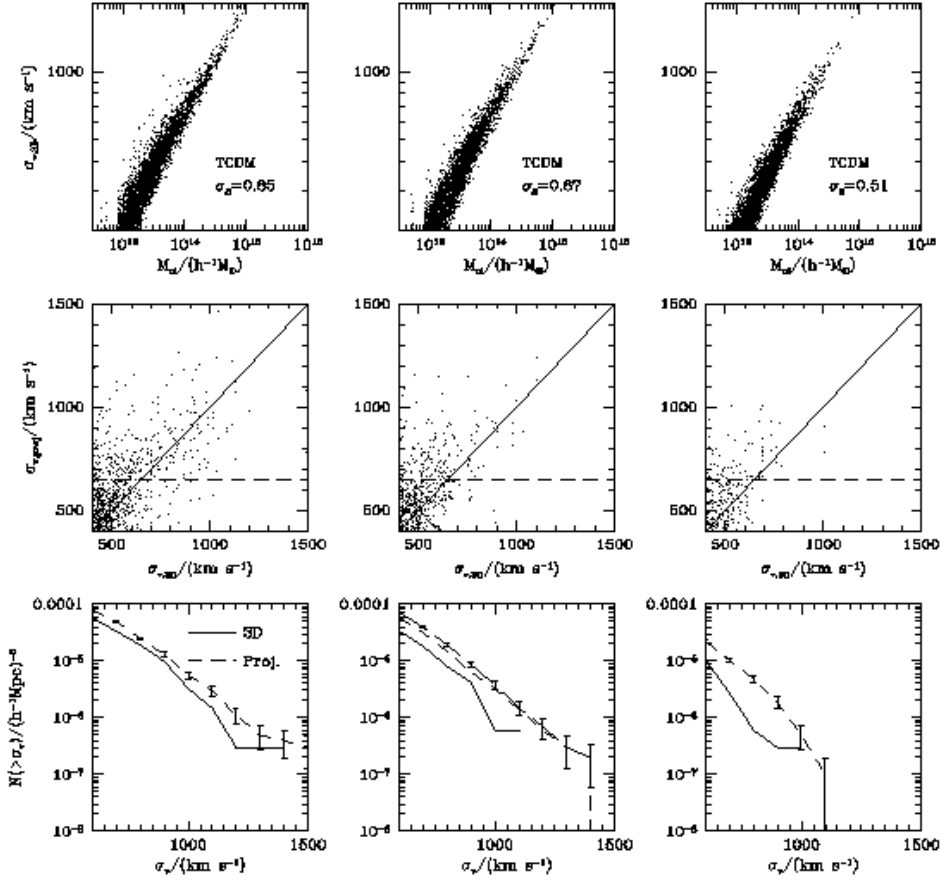}}
\caption{Effect of the observational biases on the $\sigma_v$ 
estimates for the three TCDM outputs. Panels in the first line show the 
relation between 
intrinsic velocity dispersion and cluster masses, both estimated at
$1.5\hm$. Panels in the second line show the scatter between intrinsic and 
observed cluster velocity dispersions. The horizontal dashed line is for
the $\sigma_v$ completeness limit of the F96 sample, $\sigma_{v,lim}=
650\vel$. Panels in the third line are for the intrinsic (solid lines) and
observational (dashed lines) CVDF. Errorbars represent the 1$\sigma$ 
Poissonian uncertainty.}
\label{fi:compl1}
\end{figure}

Figures 3--5 are devoted to the statistical 
description of what can be understood from measurements of cluster internal 
velocity dispersion. Results are reported for all the nine simulation 
outputs that have been considered (for each model, plotted results refer 
to only one realization). The panels in 
the first line show the relation between the intrinsic $\sigma_{v,3D}$, 
estimated by using all the simulation particles within $1.5\hm$, and the total
cluster mass within the same radius (as for CHDM, both quantities refer only
to the cold particles). We note that a well defined correlation always exists,
which reflects the condition of virial equilibrium characterizing most of the
clusters. Few clusters detach significantly from this correlation and
all have velocity dispersions larger than the virialization one. Such
exceptions are even more rare for those models, like the OCDM ones, in which
the cluster particles have spent more time within the collapsed structure and,
therefore, have had more time to reach virial equilibrium.

Note that a scatter of a factor two in mass at a fixed $\sigma_v$ value is not
rare, especially at low $\sigma_v$ values.
However, one would conclude that in general rather reliable 
cluster mass determinations would be possible from measurements of the 
internal velocity dispersion. The situation is rather different 
in realistic cases, when observational limitations are introduced. 
This is shown in the panels of the second line,
where we plot the relation between the projected $\sigma_v$, estimated 
according to the procedure described in Section 3, and the intrinsic 
$\sigma_v$. Although most of the clusters show discrepancies of about 
100--200$\vel$ between observed and true velocity dispersions, 
there are cases in which the difference is as large as $\sim 500\vel$. 

\begin{figure}
\mbox{\epsfxsize=14 cm\epsffile{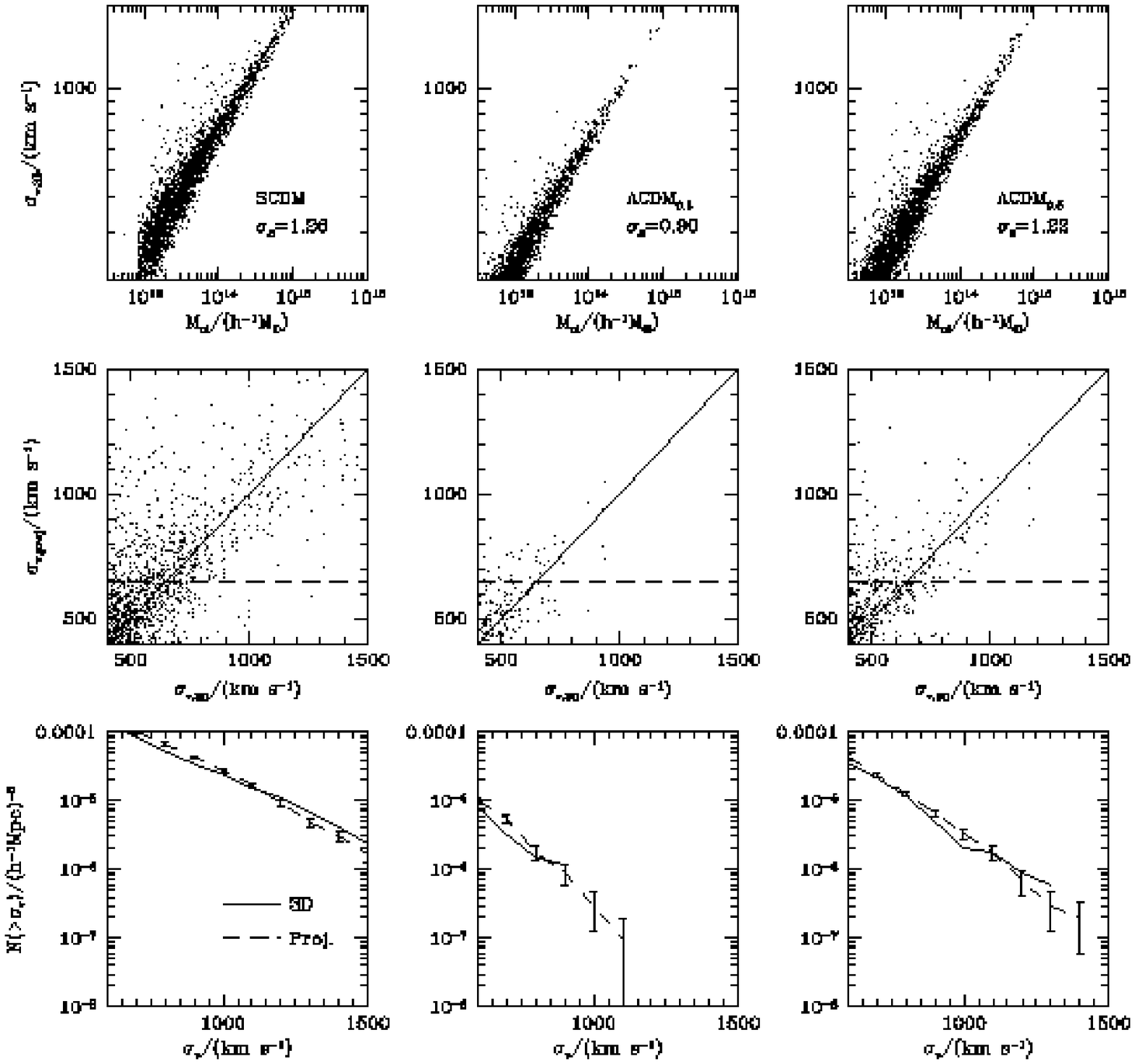}}
\caption{The same as in Figure 3, but for SCDM, $\Lambda$CDM$_{0.3}$ and
$\Lambda$CDM$_{0.5}$ simulations.}
\label{fi:compl2}
\end{figure}

The consequence of the large scatter in the $\sigma_{v,proj}$--$\sigma_{v,3D}$
on $N(>\sigma_v)$ is plotted in the panels of the third line, where we
compare the CVDF for the intrinsic (solid lines) and observed (dashed lines)
velocity dispersions. Errorbars, which are reported only for the 
``observational'' CVDF, represent the Poissonian uncertainties.
As expected, the overall effect is that of increasing the CVDF, especially 
in the large $\sigma_v$ tail. This result is qualitatively
similar to that found by van Haarlem et al. (1997). A closer comparison with his 
analysis is however rather difficult, since he considered simulations only
of the SCDM model with $\sigma_8=0.63$, by adopting a lower mass resolution,
although within a larger simulation box. Furthermore, he also used 
different procedures to remove observational biases in the $\sigma_v$ 
estimate.

We point out that all the results we reported here are based on 
clusters whose $3D$ velocity dispersion is larger than
$\sigma_{lim,3D}=400\vel$. Since we require that simulated cluster samples 
be complete above $\sigma_{lim,proj}=650\vel$ (indicated with the
horizontal dashed line), we should keep $\sigma_{lim,3D}$ small enough
that no clusters below this limit would have their $\sigma_{v,proj}$ 
increased above the observational completeness limit. On the other hand, 
the number of
selected clusters rapidly increases as $\sigma_{lim,3D}$ decreases, so as
to hardly keep the amount of data to be analysed to a manageable size.
We verified that the incompleteness induced in the simulated
samples by taking $\sigma_{lim,3D}=400\vel$ is always negligible (note that
only very few clusters have $\sigma_{v,proj}>650\vel$ at the smallest 
$\sigma_{v,3D}$). Indeed, for the TCDM model with $\sigma_8=0.67$ we plot 
the observed $N(>\sigma_v)$ obtained from $\sigma_{lim,3D}=400\vel$
(short--dashed curve) and $\sigma_{lim,3D}=200\vel$ (long--dashed curve). 
The small difference between these two cases confirms the reliability of 
cutting at $\sigma_{lim,3D}=400\vel$. 

Note that the difference with respect to the intrinsic $N(>\sigma_v)$ is 
rather model dependent being in general smaller for those models whose 
clusters have had more time to virialize; an excellent recovering of the 
3D cumulative distribution is indeed found for SCDM and $\Lambda$CDM$_{0.5}$,
which have the largest $\sigma_8$ values; instead, larger 
differences exist for TCDM as lower $\sigma_8$ values are considered. 
Also note how models whose intrinsic $N(>\sigma_v)$ are comparable, like
$\Lambda$CDM$_{0.3}$ and TCDM with $\sigma_8=0.51$ are affected in a rather 
different way by observational biases. The CVDF for $\Lambda$CDM$_{0.3}$ is
almost unaffected at $\sigma_v\mincir 900\vel$, while it only acquires a high
$\sigma_v$ tail up to $\sigma_v\simeq 1100\vel$. On the other hand, the
low--$\sigma_8$ TCDM significantly increases its CVDF over the whole 
$\sigma_v$ range. This is due to the fact that $\Lambda$CDM$_{0.3}$ 
clusters are small and rather
isolated structures, which already undergone virialization. As a consequence, 
the effect of interlopers is in general 
smaller than for TCDM clusters, which are
expected to be more extended objects (like in any $\Omega_0=1$ model) 
characterized by the presence of substructures and continuous 
merging of surrounding  clumps. This is explicitely shown in 
Figure 6, where we plot
the structure and velocity dispersion profiles for two clusters extracted from
a $\Lambda$CDM$_{0.3}$ and a TCDM ($\sigma_8=0.51$) simulation. Left and
central panels show in projection the whole particle distribution 
within the observational cylinder, before and after the inclusions of the 
observational biases, respectively. Since the two simulations are started 
with the same initial random numbers and the two clusters lie almost at the
same position, they are originated by the same waves and any difference in 
their morphology is only due to the difference in the 
corresponding cosmologies.
The cluster in $\Lambda$CDM$_{0.3}$ has a much better defined  
shape than that in TCDM. The latter, even after the interloper removal, does
not look in projection like a well defined structure. As a consequence, the
membership assignment is not very efficient in identifying genuine
cluster members and the resulting velocity dispersion profile does not recover
the true $\sigma_v$ so efficiently like for the $\Lambda$CDM$_{0.3}$ cluster.

\begin{figure}
\mbox{\epsfxsize=14 cm\epsffile{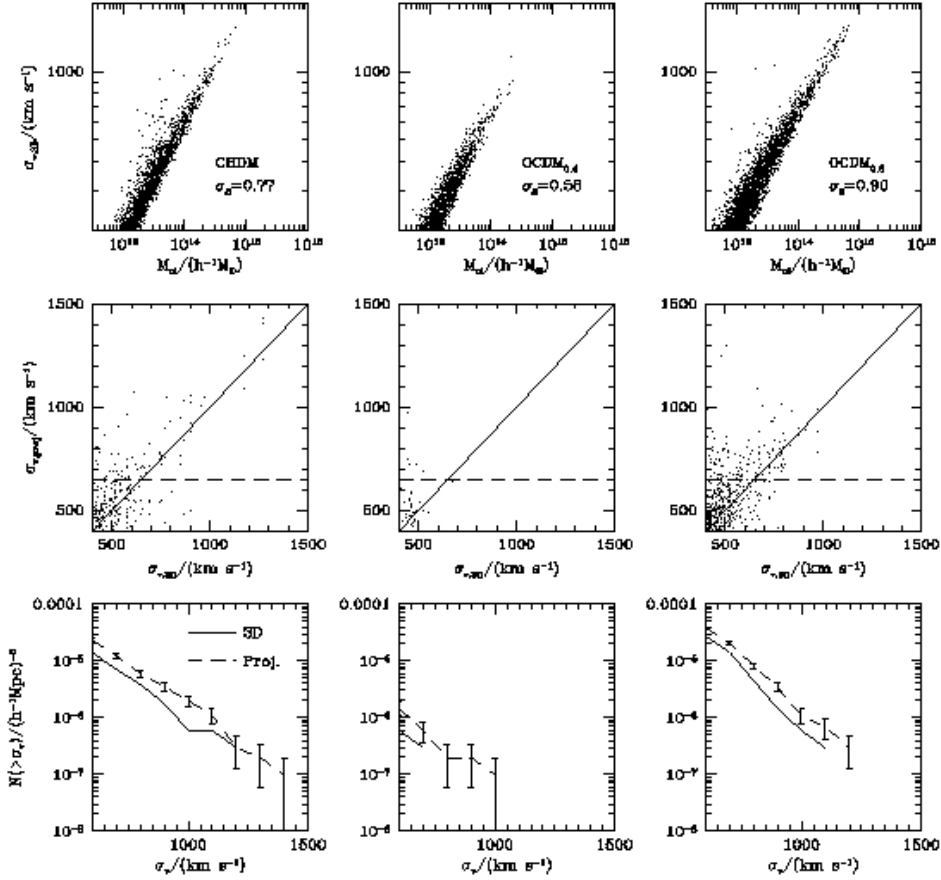}}
\caption{The same as in Figure 3, but for CHDM, OCDM$_{0.4}$ and 
OCDM$_{0.6}$ simulations.}
\label{fi:compl3}
\end{figure}

The results of this analysis can be summarized into two main points.

\begin{description}
\item[(a)] Although the scatter between the intrinsic and the observed 
$\sigma_v$ is in general rather large, the difference between 
the corresponding $N(>\sigma_v)$
is not dramatic, especially for $\sigma_v \mincir 800 \vel$. This is just the 
consequence of the roughly symmetric distribution of overestimates and
underestimates of $\sigma_v$. Discrepancies in the high $\sigma_v$ tail
are due to few clusters whose $\sigma_v$ ought to be overestimated by
200--300$\vel$, because of  the presence of unremoved interlopers. A typical
example of this occurrence is just provided by the third cluster in
Figure 2.
\item[(b)] Differences between intrinsic and observed $N(>\sigma_v)$ may
however not be negligible. Furthermore, no {\em a priori} recipe 
exists, which could allow to recover the correct CVDF from the observed one, 
the difference being non--trivially model dependent. This casts doubts on
the reliability of detailed comparisons between the abundance of clusters, as 
inferred from their internal velocity dispersion, and predictions of DM models
based on analytical approaches, like the Press \& Schechter (1974) one, which
can not include any observational effect.
\end{description} 

\subsection{Comparing with DM models}
Based on the results obtained in the previous sections, we compare in Figure
7 the CVDF for the observational sample and for the simulations. 
In each panel, the shaded band represents the result of our analysis
of the F96 sample (cf. Figure 
1 and Section 4.1). For each model, results are obtained by 
averaging over the three available realizations (exept for CHDM). 

As for the SCDM model, as expected it largely overproduces clusters at any
$\sigma_v$. In order to check for the effect of a possible velocity bias on
this model, we resorted to the relation $b_v=0.7(r/0.5\hm)^{0.2}$ ($b_v$:
ratio between the velocity dispersions of ``galaxies'' and dark 
matter particles) obtained by Evrard et al. (1994) from
hydrodynamical cluster simulations of a CDM model with $\sigma_8=0.6$.
Since these authors found the above relation to be almost independent of the
evolutionary stage, we adopt it also for our larger $\sigma_8$ 
output (cf. Table 1). The effect of such a velocity bias is shown with 
the dotted curve in the upper left panel; even in 
this case the resulting $N(>\sigma_v)$ is much larger than the 
observational one.

\begin{figure}
\mbox{\epsfxsize=14 cm\epsffile{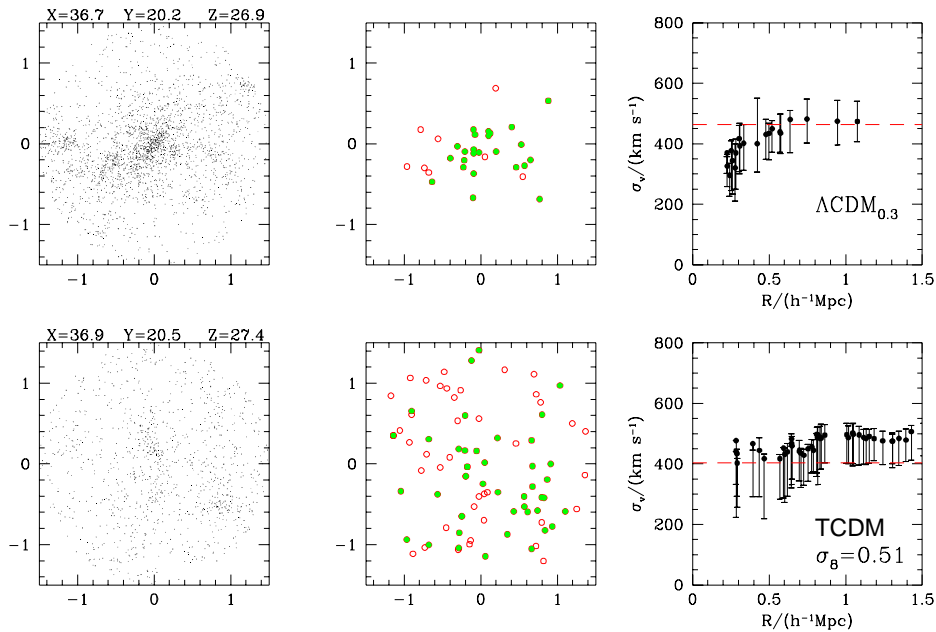}}
\vspace{-5.0truecm}
\caption{Comparison of the role of projection effects on a cluster 
identified in $\Lambda$CDM$_{0.3}$ (upper panels) and TCDM with 
$\sigma_8=0.51$ (lower panels) simulations. Note that the two clusters have 
almost the same positions in the two simulations based on the same set of 
initial random numbers. Left panels show in projection the whole distribution
of simulation particles within the observational cylinder. Central and right 
panels are the same as for plots in the first and fourth lines of Fig. 2.}
\label{fi:tlcdm}
\end{figure}

From the results reported in lower left panel, it turns out that
lowering the primordial spectral index to $n=0.85$ is not enough to
bring a $\Omega_0=1$ CDM model into agreement with observations,
unless one introduces a substantial amount of gravitational wave
contribution to the CMB anisotropy.  Even taking $T/S=7(1-n)$ from
power--law inflation ($\sigma_8=0.67$), the CVDF remains significantly
larger than that of the observational sample.  Only further lowering
the normalization to $\sigma_8=0.51$ leads into agreement with real
data, thus in accordance with the X--ray based results (see, e.g.,
E96; Pen 1996). In their analysis of the large--scale velocity fields,
Zaroubi et al. (1997) found that TCDM with $n\simeq 0.85$ is the best
fit for $\Omega_0=1$, $h=0.5$ CDM models only if $T/S=0$ (see,
however, Borgani et al. 1997b). A possible way out may be allowing for
a substantially larger baryon fraction (see, e.g., White et al. 1996).
This would have the effect of suppressing fluctuations on the cluster
mass scale, while leaving the spectrum almost unchanged at the larger
scales probed by velocity fields.

As for the CHDM model, it turns out to marginally overproduce clusters. 
A better agreements can be achieved either by increasing the hot component
or by tilting the primordial spectral index. In both cases this model, which
is already marginally consistent with observations of high--redshift 
($z=3$--4) damped Ly$\alpha$ systems, would have even harder time with 
the galaxy formation timing (see, e.g., Borgani et al. 1996, 
and references therein).
A way to overcome this problem would be to share the hot component
between more than one $\nu$ species;
in this case, the increase in the 
neutrino free streaming suppress fluctuations on the cluster mass scale, 
without significantly affecting the fluctuation power on the galaxy mass 
scale. 
A model with $\Omega_\nu =0.2$ and two massive neutrinos (Primack et
al. 1995; Primack 1996) has been found to satisy at the same time the
constraints from the cluster abundance and high--redshift objects.

\begin{figure}
\mbox{\epsfxsize=14 cm\epsffile{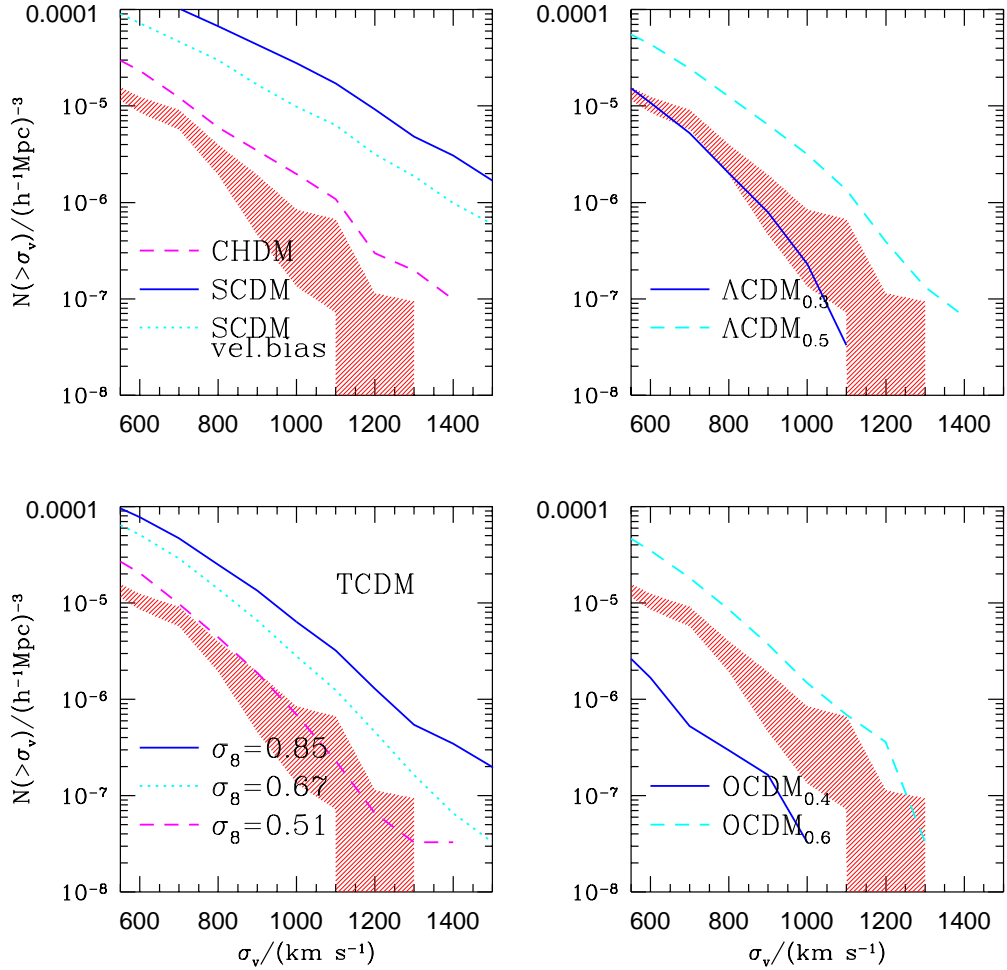}}
\vspace{-0.5truecm}
\caption{Comparisons between CVDF's for real data (dashed band) and
simulations. Simulation results correspond to the average over the three
random realization of each model (except for CHDM, for which only one 
realization is available).}
\label{fi:dmobs}
\end{figure}

In any case, it is interesting to note that the CHDM model, for which
$\sigma_8=0.77$, has a CVDF which is comparable to (or even smaller
than) that of TCDM with $\sigma_8=0.67$, even though $\Omega_0=1$ for
both models. This is the consequence of the presence of the neutrino
component, which acts like a sort of background superimposed on the
more clustered cold component. Its effect is that of slowing down the
dynamical clustering evolution, so that more time is required for a
CHDM model to develop the same small--scale velocities of a purely
cold model. This means that a larger $\sigma_8$ value is required for
a CHDM model to produce the same cluster abundance as an $\Omega_0=1$
CDM model.

As for the $\Lambda$CDM models, the $\Omega_0=0.3$ case is quite
consistent with the observational $N(>\sigma_v)$, thus confirming the
result based on $X$--ray data about the capability of this model to
produce the correct cluster abundance (see, e.g., E96).  The
$\Lambda$CDM$_{0.5}$ model, instead, turns out to overproduce
clusters, as expected on the ground of its large $\sigma_8$ value.  As
for OCDM, the $\Omega_0=0.4$ model is found to underproduce clusters,
while increasing the density parameter to $\Omega_0=0.6$ turns into a
marginal cluster overproduction.

\section{Conclusions}
In this paper we addressed the question concerning the reliability of
the distribution of cluster internal velocity dispersions as a
diagnostic for cosmological models. In order to properly address this
issue we decided: {\em (a)} to run simulations for a variety of
models, so as to verity the discriminative power of $N(>\sigma_v)$;
{\em (b)} to analyse simulated clusters in the same way as a
reference observational sample, originally considered by 
Fadda et al. (1996; F96), consisting of 153 clusters and complete in
velocity dispersion down to $\sigma_{v,lim}\simeq 650\vel$.

Our main results can be summarized as follows.

\begin{description}
\item[(a)] Projection effects and the limited number of galaxy
  redshifts per cluster can heavily pollute the recovering of the
  correct $\sigma_v$.  The cluster member selection procedure leads to
  the wrong identification of several interlopers as genuine members.
  This effect is more pronounced for low--$\sigma_v$ ($\mincir
  600\vel$) objects. On the other hand, high--$\sigma_v$ clusters,
  which look in projection like better defined structures, display an
  improved recovering of the true $\sigma_v$.

\item[(b)] The resulting effect is in general that of overestimating
  the CVDF, especially in the high--$\sigma_v$ tail. This is due to
  few small, low--$\sigma_v$ clusters, whose velocity dispersion is
  heavily boosted by projection effects. Furthermore, the amount of
  the overestimate is non--trivially model dependent: simulations in
  which particles last more within virialized structures generate
  clusters with a sharper profile, even after projection, so as to
  make interloper removal an easier task.

\item[(c)] As for the comparison with data, our results substantially
  agree with the $\sigma_8$--$\Omega_0$ scaling for CDM models, based
  on the Press--Schechter approach to the $X$--ray temperature
  function (e.g., Viana \& Liddle 1996; Eke et al. 1996; Pen 1996). We
  confirm that a $\Omega_0=1$ CDM model requires $\sigma_8\simeq 0.5$.
  The $\Lambda$CDM model with $\Omega_0=0.3$ produces the correct
  $N(>\sigma_v)$ with $\sigma_8=1.1$. This value is somewhat larger
  than that predicted by Eke et al. (1996; cf. Table 1), but rather
  consistent with the scaling provided by Pen (1996).  As for the CHDM
  model, it overproduces clusters, although not by a large amount.  In
  any case, the presence of the neutrino background has the effects of
  slowing down the development of non--linear structures. As a
  consequence, CHDM models are allowed to have a larger $\sigma_8$
  value to provide the same cluster abundance as an $\Omega_0=1$ CDM
  model (see also Jing \& Fang 1994 and Walter \& Klypin 1996). This
  is the reason why the CVDF for CHDM with $\sigma_8=0.77$ is smaller
  than that of TCDM with $\sigma_8=0.67$.
  
  Such constraints on $\sigma_8$ from cluster abundance can be
  compared with those from large--scale velocity fields. For instance,
%
  Zaroubi et al. (1997) found for CDM models that
  $\sigma_8\Omega_0^{0.6} =0.88\pm 0.15$ ($90\%$ c.l.) is predicted by
  a maximum likelyhood approach to the peculiar velocities of the Mark
  III sample.
%
  Instead, Borgani et al. (1997b) analysed the sample of cluster
  peculiar velocities by Giovanelli et al. (1997) and derived
  $\sigma_8\Omega_0^{0.6}=0.5\pm 0.2$ ($90\%$ c.l.), thus in better
  agreement with cluser abundance.
\end{description}

Based on such results, we conclude that cluster internal velocity
dispersions represent a stringent cosmological constraint, provided
that observational biases, like projection and sampling effects, are
carefully accounted for (cf. also van Haarlem et al. 1997). It is also
clear that further advancements both from the theoretical and the
observational sides would be required in order to fully exploit the
potential of cluster velocity dispersions. For instance, the
availability of high--resolution simulations, also including gas
dynamics, should settle the question of the velocity bias, whose
amount should be rather model dependent.

Furthermore, improved observational strategies should eliminate some
of the ambiguities, which are still present in the data sets. The
possibility of measuring many more galaxy redshifts per cluster will
allow a better understanding of the internal cluster dynamics and of
the role of substructures.  Extensions of the today available samples
in the following two directions would provide even more stringent
constraints on DM models. From the one hand, realising a shallower
volume--limited ($z\mincir 0.1$) survey, including at least down to
$R=0$ clusters, would extend to smaller $\sigma_v$ the range where the
CVDF is accurately sampled. From the other hand, realising a deeper
($z\magcir 0.5$) survey of rich clusters would allow to discriminate
between those models that, even though providing the correct
$N(>\sigma_v)$ at $z\simeq 0$, have different evolutionaly patterns
(e.g., Jing \& Fang 1993; Crone \& Geller 1994).
Attempts in this direction have been already pursued with the CNOCC
cluster survey (see, e.g., Carlberg et al. 1996, and references
therein), but only for a limited number of clusters. In any case, it
is clear that as deeper cluster surveys are considered, a careful
treatment of projection biases becomes a more and more delicate issue.
Multi--object spectrographs of the new generation will be needed to
measure $\sim 10^2$ galaxy redshifts for each cluster in one shot,
thus rendering feasible in the near future a substantial enlargment of
the today available data sets.

\section*{Acknowledgment}
We wish to thank Hugh Couchman for having made available his adaptive
P3M code, Silvio A. Bonometto for providing us with the code to
estimate the CHDM transfer function, and Dario Fadda, Giuliano
Giuricin, Fabio Mardirossian, and Marino Mezzetti for providing us
with their cluster sample. We also acknowledge the {\sl Centro di
Calcolo dell'Universit\`a di Perugia}, where part of the computations
have been realized. SB acknowledge SISSA for the hospitality during
the preparation of this paper.


\section*{References}

Abell G.O., Corwin H.G., Olowin R.P., 1989, ApJS, 70, 1\\
Bahcall N.A., Cen R., 1993, ApJ, 407, L49\\
Balland C., Blanchard A., 1996, preprint\\
Bardeen J.M., Bond J.R., Kaiser N., Szalay A.S., 1986, ApJ, 304, 15\\
Bartelmann M., 1995, A\&A, 303, 643\\
Bartlett J.G., Silk J., 1993, ApJ, 407, L45\\
Beers T.C., Flynn K., Gebhardt K., 1990, AJ, 100, 32\\
Bennett C.S., et al., 1996, ApJ, 464, L1\\
Bird C.M., 1995, ApJ, 445, L81\\
Bird C.M., Mushotzki R.F., Metzler C.A., 1996, ApJ, 1995, 453, 40 \\
Biviano A., Girardi M., Giuricin G., Mardirossian F., Mezzetti M., 1993,
ApJ, 411, L13\\
Bond  J.R., Jaffe A.H., 1996, Proceedings of the XXXI Moriond meeting
``Microwave Background Anisotropies'', preprint astro--ph/9610091\\
Borgani S., da Costa L.N., Freudling W., Giovanelli R., Haynes M.,
Salzer J., Wegner G., 1997, ApJ Letters, submitted\\
Borgani S., Lucchin F., Matarrese S., Moscardini L., 1996, MNRAS, 280, 749\\
Borgani S., Moscardini L., Plionis M., G\'orski K.M., Holtzman J., Klypin A.,
Primack J.R., Smith C.C., Stompor R., 1997a, NewA, 1, 321\\
Bunn E.F., Liddle A.W., White M., 1996, Phys. Rev., D54, 5917\\
Buote D.A., Tsai J.C., 1996, ApJ, 458, 27\\
Burles S., Tytler D., 1996, Science, submitted, preprint astro--ph/9603070\\
Carlberg R.G., 1994, ApJ, 433, 468\\
Carlberg R.G., Morris S.L., Yee H.K.C., Ellingson E., 1996, ApJ
Letters, in press, preprint astro--ph/9612169\\
Cen R., 1996, preprint astro--ph/9608070\\
Collins C.A., Guzzo L., Nichol R.C., Lumsden S.L., 1995, MNRAS, 274, 1071\\
Copi C.J., Schramm D.N., Turner M.S., 1995, Phys. Rev. Lett., 75, 3981\\
Couchman H.M.P., 1991, ApJ, 268, L23\\
Crittenden R., Bond J.R., Davis R.L., Efstathiou G., Steinhardt P.J.,
1993, Phys. Rev. Lett., 71, 324\\
Crone M.M., Evrard A.E., Richstone D.O., 1996, ApJ, 467, 489\\
Crone M.M., Geller M.J., 1994, AJ, 110, 21\\
David L.P., Jones C., Forman W., 1995, ApJ, 445, 578\\
Dell'Antonio I.P., Geller M.J., Fabricant D.G., 1994, AJ, 107, 427\\
Dressler A., Schechtman S.A., 1988, AJ, 95, 985\\
Edge A.C., Stewart G.C., 1991, MNRAS, 252, 428\\
Eke V.R., Cole S., Frenk C.S., 1996, MNRAS, 282, 263 (E96)\\
Ensslin Q.A., Biermann P.L., Kronberg P.P., Wu X.P., 1996, ApJ, in
press, preprint astro--ph/9609190\\
Escalera E., Biviano A., Girardi M., Giuricin G., Mardirossian F., Mezzetti
M., 1994, ApJ, 423, 539\\
Evrard A.E., Metzler C.A, Navarro J.F., 1996, ApJ, 469, 494\\
Evrard A.E., Summers F.J., Davis M., ApJ, 1994, 422, 11\\
Fadda D., Girardi M., Giuricin G., Mardirossian F., Mezzetti
M., 1996, ApJ, 473, 670 (F96)\\
Fields B.D., Kainulainen K., Olive K.A., Thomas D., 1996, New Ast., 1, 77\\
Frenk C.S., Evrard A.E., White S.D.M., Summers F.J., 1996, ApJ, 472, 460\\
Frenk C.S., White S.D.M., Efstathiou G., Davis M., 1990, ApJ, 351, 10\\
Giovanelli R., Haynes M.P., Herter T., Vogt N.P., da Costa L.N.,
Freudling W., Wegner G., Salzer J.J., 1997, AJ, 113, 53\\
Girardi M., Biviano A., Giuricin G., Mardirossian F., Mezzetti M., 1993, 
ApJ, 404, 38\\
Girardi M., Fadda D., Giuricin G., Mardirossian F., Mezzetti M., Biviano A., 
1996, ApJ, 404, 38 (G96)\\
van Haarlem M.P., Frenk C.S., White S.D.M., 1997, MNRAS, in press,
preprint astro--ph/9701103\\
Henry J.P., Arnaud K.A., 1991, ApJ, 372, 410\\
Henry J. P., Briel U.G., 1995, ApJ, 443, L9\\
Jing Y.P., Fang L.Z., 1994, ApJ, 432, 438\\
Katgert P., Mazure A., Perea J., et al., 1996, A\&A, 310, 8\\
Katz N., Hernquist L., Weinberg D.H., 1992, ApJ, L109\\
Klypin A., Nolthenius R., Primack J.R., 1996, ApJ, in press, preprint 
astro--ph/9502062\\
Kochanek C.S., 1995, ApJ, 453, 545\\
Lacey C., Cole S., 1994, MNRAS, 271, 676\\
Lineweaver C.H., Barbosa D., Blanchard A. Bartlett J.G., 1996, A\&A,
submitted, preprint astro--ph/9610133\\
Loeb A., Mao S., 1994, ApJ, 435, L109\\
Lubin L.M., Bahcall N.A., 1993, ApJ, 415, L17\\
Lubin L.M., Cen R., Bahcall N.A., Ostriker J.P., 1996, 460, 10\\
Lumsden S.L., Nichol R.C., Collins C.A., Guzzo L., 1992, MNRAS, 258, 1\\
Ma C.P., 1996, ApJ, 471, 000, preprint astro--ph/9605198\\
Mazure A., Katgert P., den Hartog R., et al., 1996, A\&A, 310, 31\\
Merritt D., 1987, ApJ, 313, 121\\
Miralda--Escud\`e J., Babul A., 1995, ApJ, 449, 18 \\
Mohr J.J., Evrard A.E., Fabricant D.G., Geller M.J., 1995, ApJ, 447, 8\\
Navarro J.F., Frenk C.S., White S.D.M., 1995, MNRAS, 275, 720\\
Oukbir J., Blanchard A., 1996, A\&A, submitted, preprint astro--ph/9611085\\
Pen U.L., 1996, ApJ, submitted, preprint astro--ph/9610147\\
Pen U.L., Spergel D.N., Turok N., 1994, Phys Rev., D49, 692\\
Pisani A., 1993, MNRAS, 265, 706\\
Pisani A., 1996, MNRAS, 278, 697\\
Press W.H., Schechter P., 1974, ApJ, 187, 425\\
Primack J.R., Holtzman J., Klypin A., Caldwell D.O., 1995,      
Phys. Rev. Lett., 74, 2160\\
Primack, J.~R. 1996, in {\it Critical Dialogues in Cosmology},
ed. Neil Turok (World Scientific, in press), astro-ph/9610078\\
Reeves H., 1994, Rev. Mod. Phys., 66, 193\\
Seitz S., Schneider P., 1996, A\&A, 305, 383\\
Squires G., Kaiser N., 1996, ApJ, 469, 73\\
Stein P., 1996, A\&A, in press, preprint astro--ph/9606162\\
Sugiyama N., 1995, ApJS, 100, 281\\
Viana P.T.P., Liddle A.R., 1996, MNRAS, 281, 323\\
Walter C., Klypin A., 1996, ApJ, 462, 13\\
White M., Viana P.T.P., Liddle A.W., Scott D., 1996, MNRAS, 283, 107\\
White M., Scott D., 1996, Comm. Ap., 18, 289\\
White S.D.M., Efstathiou G., Frenk C.S., 1993a, MNRAS, 262, 1023\\
Wu X.P., Fang L.Z., 1997, ApJ, in press, preprint astro--ph/9701196\\
Zabludoff A.I., Huchra J.P., Geller M.J., 1990, ApJS, 74, 1\\
Zabludoff A.I., Geller M.J., Huchra J.P., Ramella M., 1993, AJ, 106, 1301\\
Zaroubi S., Dekel A., Hoffman Y., Kolatt T., 1997, ApJ, in press, 
preprint astro-ph/9610226\\

\end{document}